\documentclass[preprint]{aastex}
\usepackage{psfig,emulateapj5}

\newcommand{\kms}{km~s$^{-1}$}

\begin{document}

\slugcomment{to appear in The Astronomical Journal}

\title{The Evolutionary Status of Isolated Dwarf Irregular Galaxies \\ II.
Star Formation Histories and Gas Depletion}

\author{Liese van~Zee\altaffilmark{1}}
\setcounter{footnote}{1}
\affil{Herzberg Institute of Astrophysics, 5071 W. Saanich Rd, \\ Victoria,
   British Columbia V9E 2E7, Canada}
\email{Liese.vanZee@hia.nrc.ca}

\altaffiltext{1}{Visiting Astronomer, Kitt Peak National Observatories, 
   which is operated by the Association of Universities for Research in 
   Astronomy, Inc.\ (AURA) under a cooperative agreement with the National 
   Science Foundation.}

\begin{abstract}
The results of UBV and H$\alpha$ imaging of a large sample of isolated
dwarf irregular galaxies are interpreted in the context of composite
stellar population models.  The observed optical colors are best fit by
composite stellar populations which have had approximately constant
star formation rates for at least 10 Gyr.  The galaxies span a range
of central surface brightness, from 20.5 to 25.0 mag arcsec$^{-2}$; 
there is no correlation between surface brightness and star formation history.  
Although the current star formation rates are low, it is possible to reproduce
the observed luminosities without a major starburst episode.  The derived gas
depletion timescales are long, typically $\sim$20 Gyr.
These results indicate that dwarf irregular galaxies will be able to 
continue with their slow, but constant, star formation activity for at 
least another Hubble time. 

The sample of isolated dIs is compared to a sample of star bursting
dwarf galaxies taken from the literature.  The star bursting dwarf galaxies
have many similar properties; the main difference between these two types
of gas--rich dwarf galaxies is that the current star formation is concentrated
in the center of the star bursting systems while it is much more distributed in the
quiescent dIs.  This results in pronounced color gradients for the starbursting
dwarf galaxies, while the majority of the quiescent dwarf irregular galaxies have minor 
or non--existent color gradients.  Thus, the combination of low current star formation
rates, blue colors, and the lack of significant color gradients indicates that star formation 
percolates slowly across the disk of normal dwarf galaxies in a quasi--continuous manner.

\end{abstract}


\section{Introduction}

The dwarf irregular  class contains galaxies with widely disparate physical properties; 
dwarf irregulars are usually classified as such based on appearance (irregular morphology)
and an intrinsic low luminosity, but their luminosities 
span at least 3 orders of magnitude ($\sim$ 10$^6$ -- 10$^9$ L$_{\odot}$), and current 
star formation rates span at least 4 orders of magnitude ($\sim$ 0.001 -- 1. 
M$_{\odot}$ yr$^{-1}$).  The wide range of current star formation rates is particularly
puzzling: why do some dwarf galaxies appear to have extremely high current star formation
rates, relative to their total mass, while others do not?  Is a ``starburst'' phase common
to all dwarf galaxies, so that dwarfs which currently have lower star formation rates are in
a period of quiescence before (or after) an episode of vigorous star formation?  In other words, 
is the tendency to burst an intrinsic property of the dwarf irregular class, or a 
phenomenon associated with only a subset of the class?  If the former, then the
disparate observed properties may be merely an artifact of detecting galaxies at different phases in 
a common star formation history, while the latter would suggest that the rarer starbursting dwarf
galaxies are not representative of the class. 
In this paper, the results
of UBV imaging, H$\alpha$ imaging, and neutral hydrogen measurements are used to investigate
the star formation histories of a sample of isolated dwarf irregular galaxies which span a range of optical 
sizes, surface brightnesses, and current star formation rates, in order to address the question of 
whether the majority of normal dwarf irregular galaxies have similar star formation
histories, and to investigate whether these ``typical'' dwarf galaxies are related to the class 
of starbursting dwarfs.

The use of global optical colors as a tracer of star formation history is a well developed subject 
\citep[e.g.,][]{T80}.  Recent improvements in galaxy evolution models 
have led to the ability to model composite stellar populations with a variety of star formation
histories, stellar metallicities, and initial mass functions \citep[e.g.,][]{Le96}. 
Despite these advances, the results of galaxy evolution models for dwarf irregular galaxies are not
significantly different than those first obtained 30 years ago. 
Three conceptually different schematic star formation histories are able to reproduce the observed optical colors
of dwarf irregular galaxies:  
(1) constant star formation rates for approximately a Hubble time \citep[e.g.,][]{GHT84,HG86}; 
(2) recent elevated star formation rates superposed on an older stellar population \citep[e.g.,][]{SSB73,H77} 
and (3)  recent onset of star formation activity \citep[e.g.,][]{SS72}.  The latter is similar to 
model (2), but has no underlying old stellar component; both (2) and (3) can be summarized as ``burst'' models, 
where the current star formation rate is significantly higher than the average past star formation rate. 
As one would expect for galaxies which are generally the bluest of the normal gas--rich galaxies,
all of these models indicate that the dominant stellar populations in dwarf irregular
galaxies are ``young,'' i.e., formed within the last Gyr or so.  The underlying difference between 
these models is whether the current star formation activity is typical of the past star formation 
activity (constant star formation rate model), or if the current star formation rate is elevated over the 
past rate (burst model).  Thus, to determine which of these models best fits the ``typical'' dwarf irregular galaxy, 
it is necessary to have additional information about the evolutionary status, such as the current star 
formation rate, gas mass fraction, and stellar and gaseous metallicities.

Systematic surveys of the current star formation activity in dwarf irregular galaxies have
shown that galaxies selected on the basis of extreme blue colors (i.e., the blue compact dwarf
galaxies) have unusually high current star formation rates, and thus are best described by
bursting models \citep[e.g.,][]{SS72,MMHS97,AB98}.   In contrast, H$\alpha$ imaging of large
samples of dwarf galaxies indicates most dwarf irregular
galaxies  have extremely low star formation rates, with only a few HII regions 
distributed sparsely across the optical disk \citep[e.g.,][]{SHK91,HHG93,MH94,M96,HAB99,RH00}.  
However, the issue of whether these low star formation rates are representative of the typical star formation
activity in these galaxies remains, since low star formation rates are also expected during
the quiescent phase of an episodic starburst model.

On the theoretical side, early models of star formation activity in dwarf irregular galaxies
indicated that small disks would be unstable to star formation activity.
In particular, episodic bursts of star formation appeared to arise naturally
from stochastic self--propagating star formation (SSPSF) models \citep{GSS80}.  
However, recent work on the gas distribution in normal and low surface brightness (LSB) dwarf galaxies 
indicates that the gas density is well below the threshold for star formation across the
entire stellar disk, suggesting that it is difficult to initiate a global
burst of star formation, even in intrinsically small galaxies \citep{HP96,vZ97c,HEB98}. 
Thus, a more accurate representation of the star formation process in dwarf 
galaxies may be that of random percolation of star formation activity across the
stellar disk.  This picture is fundamentally different than classic SSPSF since the star 
formation activity does not propagate from one cell to another; rather, star formation  
occurs only in a few localized regions where the local gas density is 
sufficient to permit molecular cloud formation and subsequent star formation. 
The onset of star formation activity in this type of model is still a stochastic
process, but the star formation activity across the galaxy is expected to be more quiescent since
it is unlikely that the entire disk will have sufficient gas density to permit
a sudden starburst.  

One additional key factor is that these two models for star formation activity in dwarf
galaxies predict different radial distributions for the composite stellar populations.
Not only are the global star formation histories different (i.e., episodic starbursts vs.
quasi--continuous star formation activity), but the local star formation histories
will be different as well.  In a starburst model, a region of the stellar
disk ``turns on'' simultaneously with a very high rate of star formation while the
remainder of the disk (usually the outskirts) remains quiescent.  This centralized
star formation activity will result in a radial color gradient in starbursting
systems \citep[as observed
by  e.g.,][]{Pe96b,MMHS97,MMH99,DCPST97,DCC99}.  In contrast,  a low level of 
star formation activity
which percolates across the disk in a random manner will result in minor local color variations,
but no large scale color gradient, provided that there is no
favored location within the disk.   Thus, one
way to distinguish between a starburst model and a percolating star formation model is to
investigate the spatial distributions of the composite stellar populations in dwarf galaxies.

Ideally, star formation histories should be derived directly from the distribution of
resolved stars on a color--magnitude diagram (CMD), where the number of stars in each
evolutionary phase can be used to infer the past star formation history
\citep[e.g.,][]{TGMF91,GMTF93,MTGF95,HG95,T96,DP97,DP98,LTNH98,Ge98,Te98,GTCMLNS98,SHCG99}.  
However, most dwarf galaxies are too distant to make such an analysis feasible, even with the high spatial
resolution of the {\sl Hubble Space Telescope}.  Thus, the present study
relies on the integrated optical (UBV) colors in order to derive the star formation
histories of a large sample of  dwarf galaxies.  The observations and
data reduction of the sample are described in van Zee (2000, hereafter \citeauthor{vZ00a}).
As described in \citeauthor{vZ00a}, the sample was selected to include isolated galaxies
in order to minimize the possible effects of recent interactions on the star formation histories.

This paper is organized as follows.  A brief summary of the sample
selection and data acquisition and reduction is presented in Section 2
(see \citeauthor{vZ00a} for full details).  The observed colors are interpreted
in the context of composite stellar populations 
 in Section 3. Correlations between the current star formation rate and
global parameters, such as gas mass fraction, are discussed in Section 4. 
 Section 5 contains a brief summary of the conclusions.

\section{Observations}
The observations and data reduction for the UBV and H$\alpha$ images
were discussed in \citeauthor{vZ00a}.   A brief summary of the
sample selection and observational procedures are presented
in this section.

\subsection{Sample Selection}
The UBV imaging sample was drawn primarily from the 
{\em Uppsala General Catalog of Galaxies} \citep[UGC]{UGC}.
All of the selected galaxies were classified as late--type spirals
or irregular, with absolute magnitudes fainter than --18.0,
and with heliocentric velocities less than 3000 \kms.
Galaxies with known neighbors within 30\arcmin~and 500 \kms~were 
excluded (nearer neighbors have subsequently been 
identified for some of the objects; see \citeauthor{vZ00a} for more details).
Projected distances between the target galaxies and
their nearest known neighbor are on the order of 100 -- 200 kpc,
which translates into a crossing time of approximately 1 Gyr.
Furthermore, the nearest known neighbor is not usually an
extremely luminous system; most often, the neighboring system
is another low luminosity galaxy.  Thus, the selection criteria
result in a sample which contains low luminosity systems which 
are unlikely to have had significant interactions within the 
last several Gyr.

\subsection{Optical Imaging}
Optical images of the isolated dwarf galaxy sample were obtained with the 
Kitt Peak 0.9m telescope during 1996--1999.  Full details of the observations
and data reduction are presented in \citeauthor{vZ00a}.
Briefly, photometric images were obtained of 48 late--type
galaxies in U, B, and V; H$\alpha$ images were obtained
for 51 galaxies.  The optical images were reduced 
following standard procedures.  After sky subtraction and
masking of the foreground stars, the apparent magnitudes
and optical colors were measured.

The observed properties of the
galaxies in the UBV imaging sample are tabulated in Table \ref{tab:obs}.
This table contains: (1) UGC number; (2) an alternate name 
for the system; (3)  morphological type from the \citeauthor{RC3};
(4)  inclination angle as derived from the optical axial ratios
at the lowest surface brightness level detected in the V--band images,
assuming $q_0$ = 0.2; (5)  observed apparent B magnitude, corrected
for Galactic extinction; (6) observed \bv~ color within a 
25 mag arcsec$^{-2}$ aperture, corrected for Galactic extinction;
(7) observed \ub~ color within the same aperture, corrected for
Galactic extinction; (8) integrated HI flux density; 
(9) heliocentric velocity; (10) width of
the HI line profile, measured at 20\% of the peak; (11) derived
M$_{\rm H}$/L$_{\rm B}$; and (12) source of the tabulated HI data.
 The Galactic extinction correction was obtained from  \citet{SFD98},
and is based on the dust distribution of the Milky Way.
Since the internal extinction 
correction is uncertain for low metallicity galaxies, no
correction was applied.  The HI data were compiled from a variety
of sources \citep{FT81,TF88,Se90,Se92,GH93,Ge97,vZ97d,He98}; data from 
recent observations and small telescopes (i.e., large beams) were favored in 
the final tabulation.  In the few cases where  $W_{20}$ was not included 
in the published paper, the original HI data were obtained, and the width
was remeasured at $W_{20}$.

Distances to most of the galaxies in this sample are uncertain.
A few of the galaxies are near enough that distances have been
measured directly: 
UGC 685 -- \citet{H99}; UGCA 292 -- \citet{Me98};
UGC 9128 -- \citet{ATK00}; DDO 210 --
\citet{L99}; UGC 12613 -- \citet{Ge98}.
For the remainder, the adopted distance was calculated from
the HI recessional velocity, assuming a Hubble constant of
75 km s$^{-1}$ Mpc$^{-1}$ and using a nonlinear Virgocentric
infall model based on the outline of \citet{S80}.

Distance dependent parameters are tabulated in Table \ref{tab:derived}.
This table contains: (1) UGC number or other primary designation; (2) adopted
distance; (3) absolute magnitude; (4) extrapolated central surface
brightness for an exponential disk, corrected for Galactic extinction and inclination effects; 
(5) scale length of the exponential disk; (6) gradient in the \bv~color;
(7) gradient in the \ub~color;
(8) logarithm of the total HI mass; (9) current star formation rate;
(10) the average past star formation rate; and (11) the gas depletion time
scale.  The current star formation rate was calculated from
the observed H$\alpha$ luminosity, using the conversion factor
from \citet{K98}:
\begin{equation}
SFR =7.9 \times 10^{-42} L(H\alpha)
\end{equation}
where $SFR$ is the current star formation rate in solar masses per year
and $L(H\alpha)$ is the H$\alpha$ luminosity in erg s$^{-1}$.  The above
conversion factor assumes a \citet{S55} initial mass function (IMF)
and solar abundances, and is a slight ($\sim$12\%)
increase in the SFR relative to the SFR conversion factor of \citet{HG86}.  The average
past star formation rate was derived from the observed luminosity and the assumption
that the galaxy has been forming stars for 13 Gyr; a complete description of
the calculation of the average past star formation rate is presented 
in Section \ref{sec:life}.  The gas depletion time scale provides an indication of 
how long star formation can
continue at the present rate before exhausting the fuel supply, and was
calculated by taking the total mass of hydrogen and dividing by the current star
formation rate.

\subsection{Optical Spectroscopy of UGC 5205}

Low resolution optical spectra of UGC 5205 were obtained 
at the Palomar 5m telescope\footnote{Observations at the Palomar 
Observatory were made as part of a continuing cooperative agreement 
between Cornell University and the California Institute of Technology.} 
during the night of 1997 January 10.  The observations were undertaken in
a  manner similar to those described in \citet{vZ98}.
Two 1200 sec exposures were obtained with the long slit centered on
UGC 5205 and aligned along the parallactic angle (0\arcdeg); 
the slit was 120\arcsec~in length, and a 2\arcsec~aperture
was selected.  The spectra were acquired with matched gratings on the blue
and red sides of the Double Spectrograph, providing an effective resolution
of 9.2 \AA~(2.19 \AA/pix) for the blue camera and 7.8 \AA~(2.46 \AA/pix)
for the red camera.  The final combined spectra provided complete 
spectral coverage from 3600--7600 \AA. 
The night was non--photometric, but relative flux calibration was obtained 
by observations of spectrophotometric standard stars \citep{MSBA88}. 

\section{Star Formation Histories}
The results of the UBV and H$\alpha$ imaging observations are used in this Section to derive
approximate star formation histories. 
The global colors of the isolated dI sample are presented in Section \ref{sec:col}.
The stellar population models are described in Section \ref{sec:mods}.  The global
colors of the isolated dI sample and a comparison sample of starbursting
dwarf galaxies are used to derive average star formation histories in Section \ref{sec:sfh}.
Finally, the star formation histories based on the global
colors are checked for consistency with current and past star formation
rates in Section \ref{sec:life}.

\subsection{Global Colors}
\label{sec:col}

The analysis presented in the subsequent sections is based primarily on the observed global colors.
As mentioned in the Introduction, one of the fundamental differences between the two different models
of star formation activity in dwarf galaxies is the possibility that starbursting dwarf galaxies
will have substantial color gradients.  Since a global color will only measure the luminosity

\psfig{figure=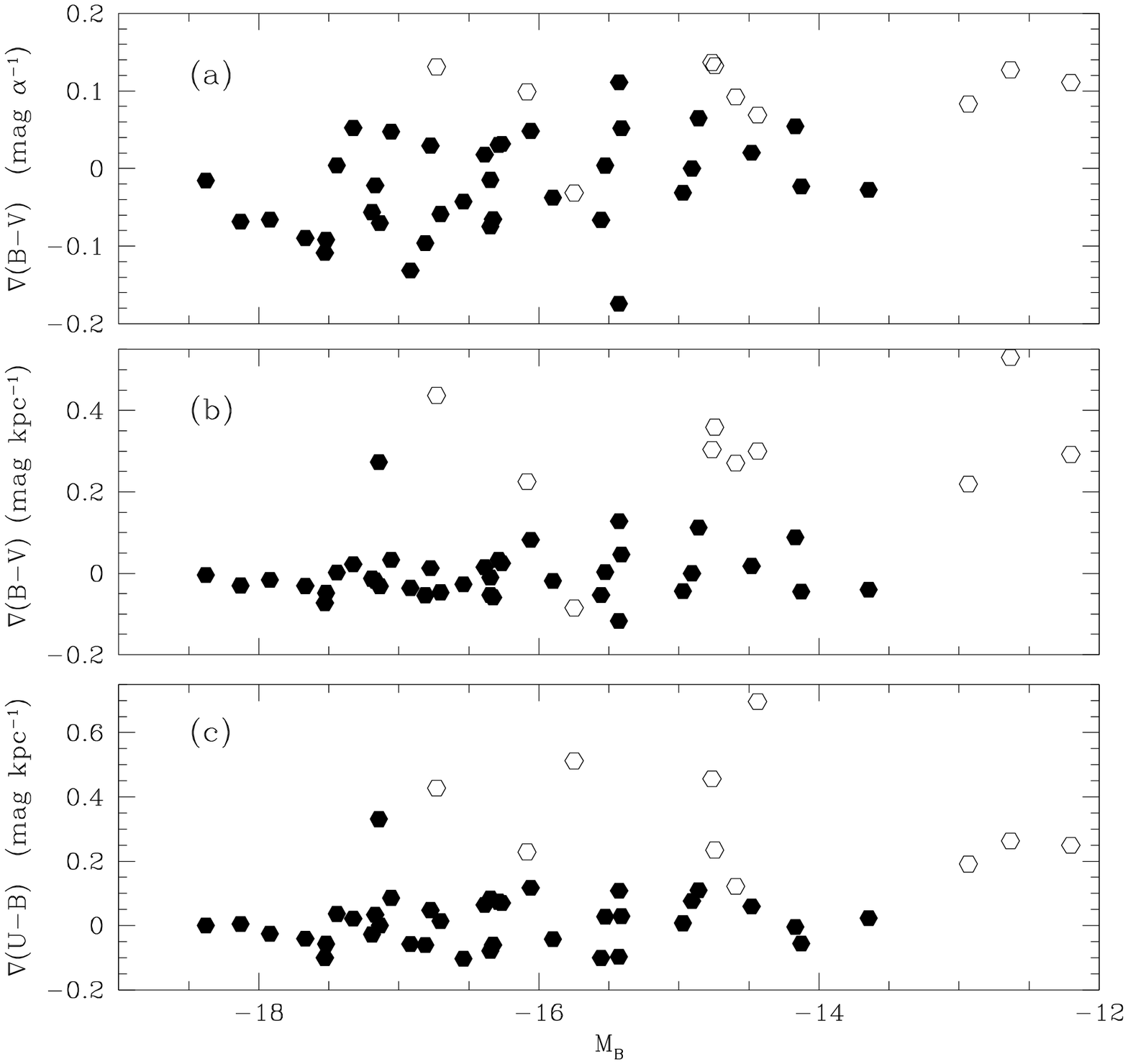,height=8.1cm,angle=0,bbllx=20 pt,bblly=170 pt,bburx=700 pt,bbury=700 pt,clip=t}
\figcaption{Observed color gradients in the dwarf irregular sample. (a) The \bv~ color
gradient in units of exponential disk scale length; (b) the \bv~ color gradient in
 units of mag kpc$^{-1}$; (c) the \ub~ color gradient in units of mag kpc$^{-1}$. 
 Galaxies with small scale lengths ($\alpha <$ 0.5 kpc) are shown with open symbols in all plots.
Most of the dIs have negligible color gradients.  However, the central regions of compact
 dIs are bluer than their outskirts.
\vskip 10pt
 \label{fig:grads}}

\noindent weighted
average of the dominant stellar population, any system with a substantial color gradient will
be misrepresented by a global color. 

 The observed color gradients for the isolated dI galaxies 
are shown in Figure \ref{fig:grads}.
The top panel shows the \bv~color gradient in terms
of exponential scale length (mag $\alpha^{-1}$).   
Interestingly, if the \bv~and \ub~color gradients are shown in terms of a physical
linear scale (mag kpc$^{-1}$; Figure \ref{fig:grads}b,c), a handful of galaxies
separate from the general trends.  These galaxies are slightly redder in their
outskirts, and, as a class, tend to have very short scale lengths for their luminosity.
Given their small sizes and color gradients, these ``compact dIs'' may be related
to the starbursting dwarf galaxies.  Nonetheless, the  net differences between
the global colors and the colors for the inner and outer regions are minor since these
galaxies are quite small; for example, assuming that the galaxian light has been traced to 
1.5 $\alpha$ (approximately 26.5 mag arcsec$^{-2}$ for this sample of galaxies), this 
corresponds to only $\sim$ 0.15 mag color difference between the inner to outer regions 
of the disk for the most extreme cases.  
Thus, it appears that the radial color gradients are negligible for the majority of galaxies in this
sample.  

The extinction corrected global colors of the isolated dwarf irregular galaxy
 sample are shown in Figure \ref{fig:ubv};
the median colors for the  sample are \bv~= 0.42 $\pm ^{0.04}_{0.05}$ and 
\ub~= --0.22 $\pm ^{0.04}_{0.07}$.  Also shown on Figure \ref{fig:ubv} are the optical colors for
two fiducial galaxies: the Large Magellanic Cloud (LMC) \citep{H78}
and HI 1225+01 \citep{SdMGH91}.   The locations of these two galaxies on the UBV color--color
diagram illustrate the extreme range of colors found in dwarf irregular galaxies.
The LMC is slightly more luminous than the class of object
considered here, but, more importantly perhaps, its star formation history has almost certainly
been influenced by interactions with

\psfig{figure=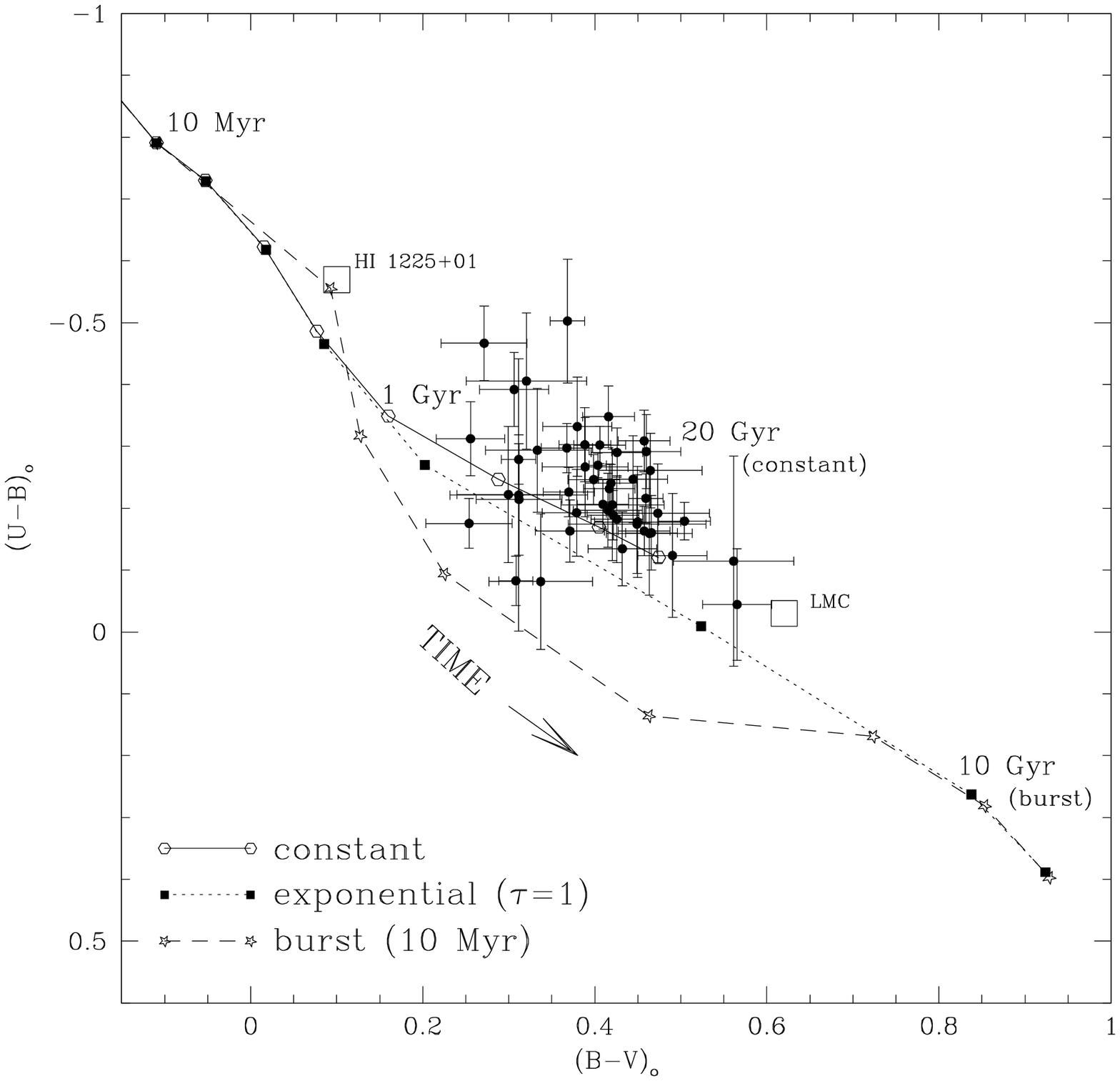,height=8.cm,angle=0,bbllx=20 pt,bblly=170 pt,bburx=700 pt,bbury=700 pt,clip=t}
\figcaption{ Color--color diagram of the isolated dwarf galaxy sample.  Three evolution
tracks for low metallicity (1/5 solar) composite stellar populations from the 
Bruzual \& Charlot (1996) models are
shown: a short (10 Myr) burst of star formation (long dashed lines); an exponentially
decreasing star formation rate, with and e--folding time of 1 Gyr (short dashed lines);
and a constant star formation rate (solid lines).  The models are marked every 0.5 dex,
and time increases from the upper left hand corner to the lower right hand corner.  The
short burst and exponentially decreasing star formation rates both result in a red
galaxy after a Hubble time, while a constant star formation rate results in a blue
system, similar to the observed properties of the dwarf irregular galaxies.  Also
shown are the global colors of HI 1225+01 (Salzer et al.\ 1991) and the LMC (Hardy 1978).
\vskip 10pt
 \label{fig:ubv}}

\noindent 
 its near neighbor, the Milky Way.  Detailed color--magnitude 
diagrams indicate that the star formation history of the LMC was somewhat ``bursty'' in nature
[e.g., \citet{vdB91,G95}, but see also \citet{He99}].  The slightly redder colors of the
LMC are naturally explained if the LMC is in a predominately quiescent phase after
a major episode of star formation.  At the other extreme, HI 1225+01 has much bluer colors than the
typical dwarf irregular galaxy, and may be an example of a galaxy dominated by a young
stellar population.  While still controversial, it has been suggested that HI 1225+01
may be an example of a galaxy at the present epoch forming stars for the first time \citep{SdMGH91}.

Despite the large differences in colors between these two fiducial dwarf irregular galaxies, 
one  remarkable aspect of the data shown in Figure \ref{fig:ubv} is that the majority of
the isolated dwarf irregular galaxies have similar UBV colors.  In fact, the entire sample
occupies only a small locus in the UBV color--color diagram.  Based on this remarkably small 
range of optical colors, it appears that the majority of isolated dwarf irregular galaxies
 have had similar star formation histories, with little dependence on total
 luminosity or surface brightness.

\subsection{Composite Stellar Population Models}
\label{sec:mods}
The goal of galaxy evolution modeling is to use the observed properties of the present
composite stellar population to infer how the star formation rate varied with 
time.  The models quickly become degenerate  since the observed 
colors provide only 
a luminosity weighted measure of the integrated light of the composite stellar populations. 
In particular, old (faint) stellar populations can be over shadowed by young (luminous) stellar 
populations; in addition, variations in either the initial mass function (IMF) or stellar
metallicity can mimic changes in the star formation rate.  Thus, in practice, it is necessary 
to determine not only the star formation history (SFH), but also how both the IMF and 
metallicity of the stellar population changed with time.  Given these degeneracies, a plethora 
of galaxy evolution models can be found to fit any combination of colors, particularly if only a few
bandpasses are observed.  However, by restricting the possible star formation histories
to simple variations of the star formation rate (an SFR which changes smoothly with time, or
one with well defined starbursts, for example), it is possible to derive reasonable SFH models 
for galaxies with a wide range of colors \citep[e.g.,][]{LT78}.

The galaxy evolution code of Bruzual \& Charlot (1996)  (\citeauthor{BC96}, hereafter)
was used to investigate color evolution tracks for a variety of possible
star formation histories and stellar metallicities.  The BC96 galaxy evolution code is an 
improved version of the composite stellar population models originally described in \citet{BC93}.  
As in \citet{BC93}, the mono--metallicity composite stellar populations are built through 
the synthesis of stellar evolutionary tracks \citep[e.g.,][]{LCB97}.   The version of the galaxy evolution
code run here includes options for sub-- and super--solar metallicity stellar populations, 
in addition to allowing the user to vary the star  formation history and the initial mass 
function.   However, the derived evolutionary tracks are not fully self--consistent, since 
neither the stellar metallicity nor the IMF are permitted to vary as a function of time.
For simplicity, a \citet{S55} IMF was adopted for all of the models considered here;
a top-- or bottom--heavy IMF will result in minor variations in the evolutionary tracks, but
relative conclusions  should be robust with regard to the choice of IMF.   
On the other hand, the lack of a self--consistent treatment of the 
chemical enrichment of subsequent stellar populations could have a severe impact on the 
derived star formation histories because of the well known age--metallicity degeneracy.  
Mono--metallicity models with similar star formation histories 
 were used to explore the possible impact of the age--metallicity
degeneracy on the derived stellar population ages (Figure \ref{fig:models}).
Each of the four panels in Figure \ref{fig:models} represent possible star formation
histories; the first three are for exponentially decreasing star formation rates
[$\Psi(t)=\tau^{-1}~{\rm exp}~(-t/\tau)$; abbreviated as ``tau models'', hereafter]
and the last is for a constant star formation rate.  In the small $\tau$ models ($\tau$ = 1 or 5),
the age--metallicity degeneracy is clearly evident in the overlapping age and metallicity
contours for red systems.  However, for blue objects (galaxies with \bv $<$ 0.5), the 
age--metallicity degeneracy is not as severe a problem;  the metallicity induced variations in 
the evolutionary tracks primarily effect the U--B colors in these models, and have a small
amplitude.    Since the observed colors of the isolated dI sample
are all quite blue (\bv $<$ 0.5),  the derived composite stellar population models should
provide reasonably accurate ages for the dominant stellar populations. 

The derived galaxy evolution models included those for 

\psfig{figure=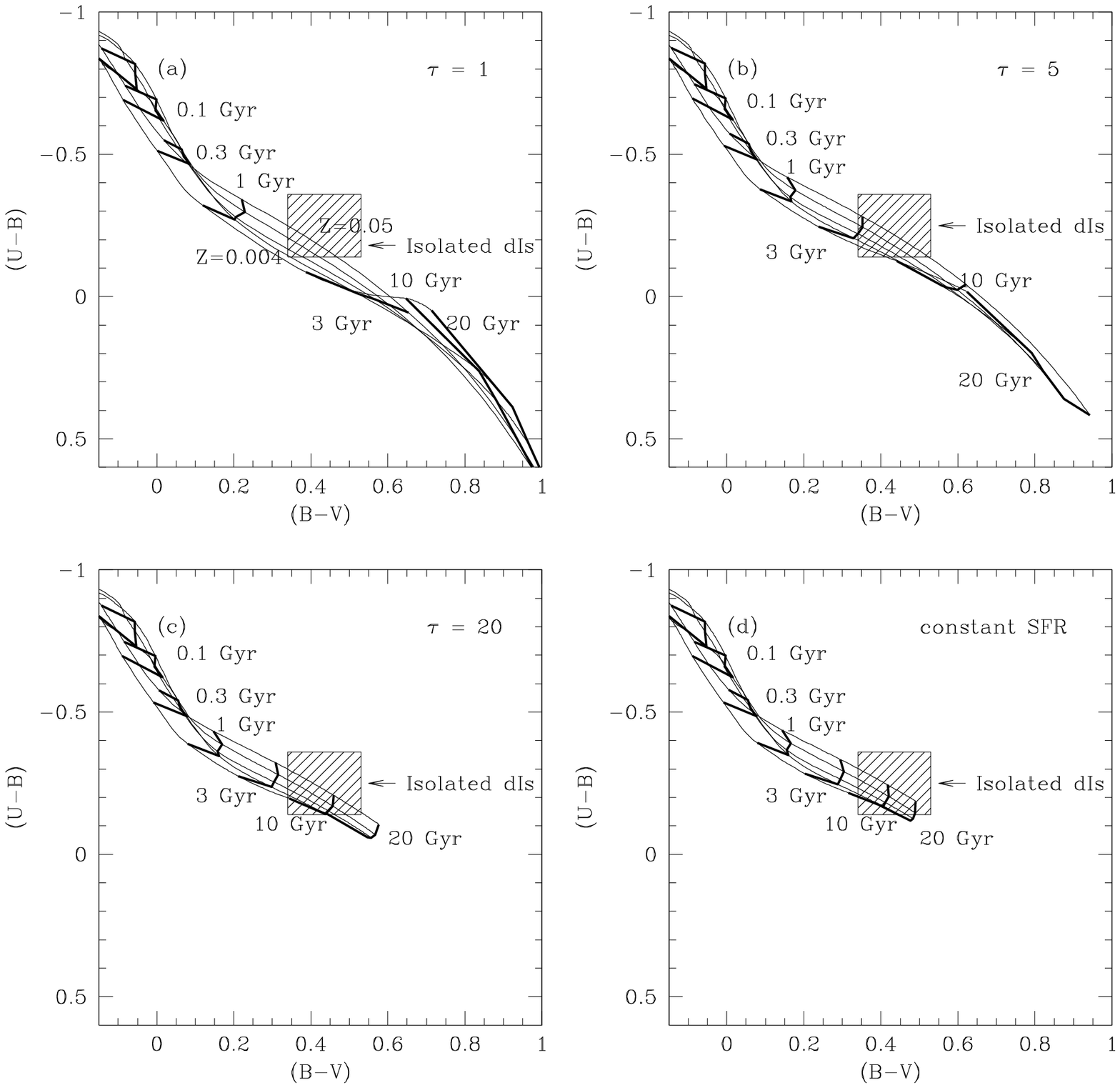,height=8.cm,angle=0,bbllx=20 pt,bblly=160 pt,bburx=700 pt,bbury=700 pt,clip=t}
\figcaption{ Color--color diagrams for mono--metallicity composite stellar populations
from the Bruzual \& Charlot (1996) stellar population models.  Each panel represents
a different star formation history, and shows the evolutionary tracks for
different metallicities (1/5, 1/2.5, 1, and 2.5 times solar). 
The shaded box denotes the locus of colors for the majority of the galaxies in the isolated dI sample. 
(a) Predicted colors for $\tau = 1$ Gyr models.  The bold lines denote the locus for specific
time steps; solid lines denote the evolutionary tracks for different metallicities.
 (b) Predicted colors for $\tau = 5$ Gyr models. (c) Predicted colors for $\tau = 20$ Gyr models. (d)
 Predicted colors for constant star formation rate models.  In all of these models, 
the age--metallicity degeneracy is less significant for blue systems (\bv $<$ 0.5).
\vskip 10pt
 \label{fig:models}}

\noindent
a single short starburst (10 Myr), a 
constant star formation rate, and several variations of exponentially decreasing star formation 
rates.  At minimum, all star formation history models were run for both solar and 1/5 of solar metallicity, 
and many were run for all available metallicities (1/50 to 5 times solar). 
Examples of the color evolution of several of these models are shown in Figure \ref{fig:ubv};
the color evolution tracks are marked every 0.5 dex, and time increases from the upper
left hand corner to the lower right hand corner. 
For clarity, only the 1/5 of solar metallicity evolution tracks are shown \citep[dwarf irregular
galaxies typically have gas--phase abundances of 1/50 to 1/5 of solar,][]{SKH89}.  
Galaxies with short tau parameters, or galaxies which
are dominated by a single burst of star formation, will have evolved into extremely red
galaxies by the present epoch.  In contrast, galaxies with long tau parameters ($\tau \sim$ 10--20 Gyr),
or constant star formation rates, will still be blue systems, even after a Hubble time.

In addition to models with well defined star formation histories, several more complicated 
histories were also investigated.  Figure \ref{fig:burstmodels} shows one such model, where
a recent star burst has been superposed on top of a constant star formation rate.
This particular model is for a starburst superposed on a galaxy which has had 
continuous star formation activity for 10 Gyr; the starburst consists of a brief period of time (500 Myr)
where the star formation rate is 10 times the average past rate; after the starburst episode,
the galaxy returns to its original star formation rate.
As expected, the color evolution of the more complicated star

\psfig{figure=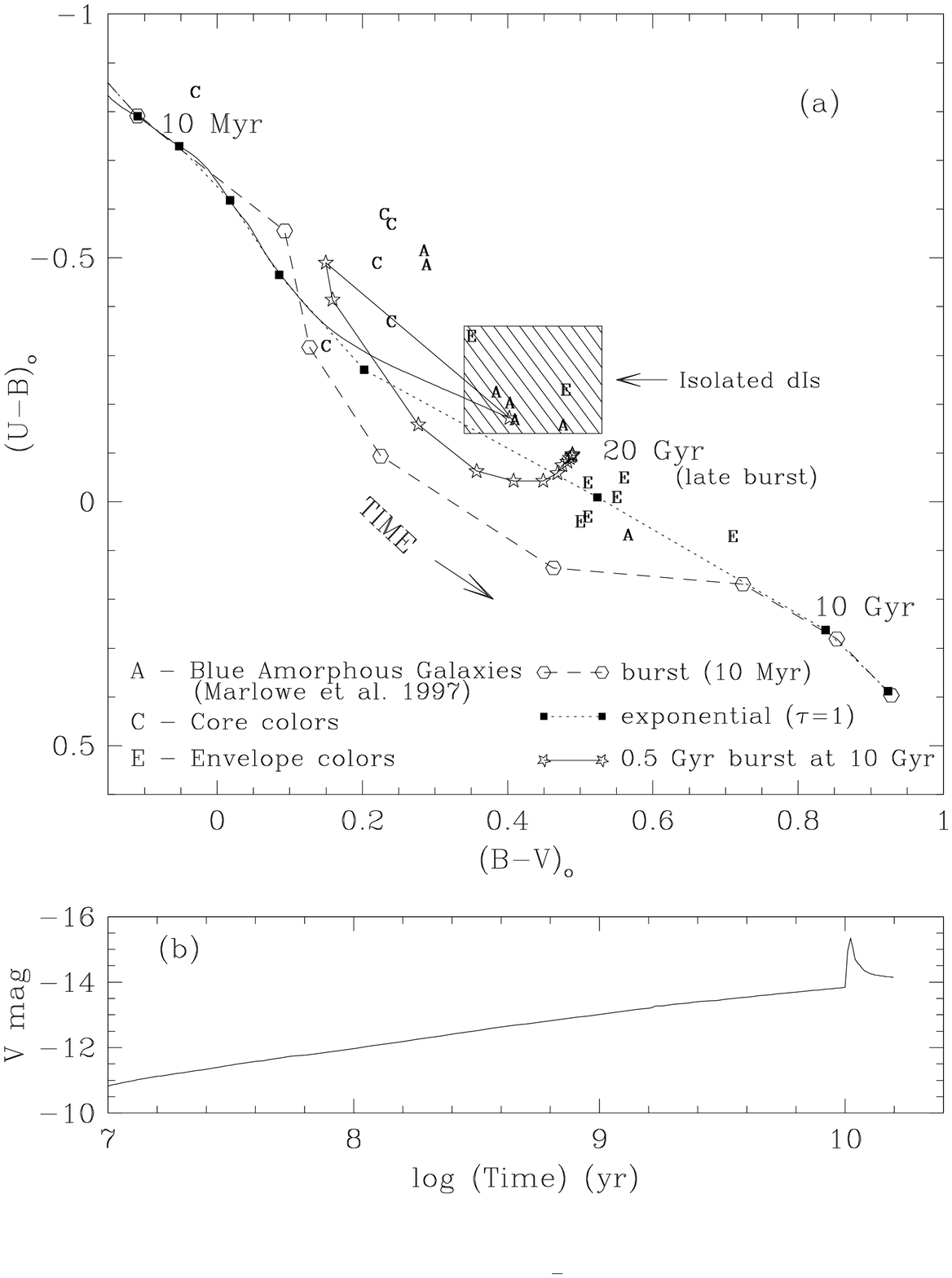,height=9.cm,angle=0,bbllx=0 pt,bblly=50 pt,bburx=700 pt,bbury=750 pt,clip=t}
\figcaption{(a) Evolutionary tracks for low metallicity (1/5 solar)
composite stellar populations from
the Bruzual \& Charlot (1996) stellar population models.   The short burst of
star formation and exponentially decreasing star formation rate are the same
as in Figure 2.  Solid lines denote the color evolution of a system which
undergoes a 0.5 Gyr burst of star formation (with  $<SFR>_{\rm burst}$ = 10 $\times <SFR>_{\rm past}$)
after having a quasi--continuous SFR
for 10 Gyr.  The stars mark time steps of 0.01 dex.  Galaxies with  \bv~ = 0.36
and \ub~ = $-0.06$ correspond to 750 Myr post--burst systems.  The shaded box denotes
the locus of colors for the majority of the galaxies in the isolated dI sample.  In addition, 
colors of blue amorphous dwarf galaxies are shown  (Marlowe et al. 1997;1999); the global colors
 of these starbursting systems are denoted by ``A'', while the colors of the central core and 
extended stellar envelopes of the same galaxies are denoted by 
``C'' and ``E'', respectively.  (b) Luminosity evolution
for a system with a short burst of star formation at 10 Gyr superposed on
a quasi--continuous star formation rate. \vskip 10pt
\label{fig:burstmodels}}

\noindent    formation histories
are a mixture of quiescent tracks and  starburst tracks. 
For the brief period of time while a galaxy is undergoing a substantial starburst, both
the \bv~and \ub~colors will be significantly affected by the young stellar population, and
the observed colors will be bluer in \ub~than expected for a simple composite stellar population.
The extent to which this occurs depends on the burst strength 
\citep[see, e.g.,][]{H77,LT78,AT90,KFLF91,vdH00}.
However, after the starburst fades, all models which combine a starburst superposed on a constant
or decreasing star formation rate result in composite stellar populations which evolve through the locus of
points between the short starburst (10 Myr, single population), and the exponentially decreasing
star formation rate models (as illustrated in Figure \ref{fig:burstmodels}).  
Thus, {\em the observed \ub~color is an excellent tracer of previous starburst
activity since any galaxy which has undergone a substantial starburst will be redder in 
\ub~than expected for a  simple composite stellar population.}  
Depending on the burst strength, these color offsets may exist for several Gyr,
as the burst population evolves through the Red Giant Branch. 
In essence, the color offset from a nominal constant star formation rate evolutionary
track is the result of the impact of the aging burst population on the luminosity 
weighted colors.  The \ub~color is particularly sensitive to an aging burst population
because is  dominated by the flux of the youngest, bluest stars; as these stars evolve, 
the \ub~color changes rapidly, while the redder colors (i.e., \bv~and \vr) change more slowly.

In the example shown here, a galaxy with \bv~= 0.36 and \ub~= $-0.06$ corresponds
to a post-burst system where the last major episode of star formation occurred within the
last 750 Myr.   Of course, there are still ambiguities in the derived ages for galaxies
in this region of the UBV color--color diagram since it is not immediately obvious if a galaxy 
is on a post--burst track (and, therefore, has an older underlying stellar population)
or if a galaxy has an intrinsically short tau parameter.  In other words, it is  
difficult to distinguish between an evolving post--burst population 
and a truly young galaxy  based on optical colors alone.   Nonetheless, these models suggest that 
galaxies which fill in this region of the UBV color--color diagram are likely to be
post--burst systems whose age (time since most recent substantial star formation episode)
can be approximated by the age of the corresponding stellar population from a short tau model.

The expected luminosity evolution of a galaxy which undergoes a brief starburst is shown in Figure 
\ref{fig:burstmodels}b.   The luminosity of the
galaxy increases smoothly during the first 10 Gyr while the galaxy has a constant star formation
rate; the starburst momentarily increases the luminosity
by approximately 1.5 magnitudes; the integrated luminosity decreases immediately following the 
starburst as the dominant stellar population ages.  Eventually, the newer stellar generations 
(formed at the lower rate) will again dominate the global colors (Figure \ref{fig:burstmodels}a) 
and the integrated luminosity (Figure \ref{fig:burstmodels}b).   If the galaxy does not undergo 
additional starbursts, the net effect will be
a slightly enhanced luminosity (relative to a constant star formation rate model)
and \ub~colors which are slightly redder than other galaxies with similar \bv~colors.

\subsection{Star Formation Histories from Global Colors}
\label{sec:sfh}
\subsubsection{The Isolated Dwarf Irregular Sample}

Comparison of the observed colors with the galaxy evolution tracks 
indicates that the majority of the dwarf irregular galaxies appear
to lie slightly above the constant star formation rate track (Figure \ref{fig:ubv}).  That is, the 
observed \ub~color is slightly bluer than predicted by the composite stellar
population models.  Constant star formation rate models with  higher metallicity
stellar populations (solar metallicity and above) provide a slightly better
fit than the lower metallicity models (see, e.g., Figure \ref{fig:models}).  
However, it is unlikely that the stellar metallicities are solar or 
super-solar, given that the gas phase metallicities of these galaxies are quite low, 
typically $\sim$ 1/10 -- 1/5 solar \citep{vZH00}.  While this discrepancy is troubling, it is 
likely an artifact of the models, and in particular, may be due to the
poorly understood physics of low metallicity massive stars
\citep[e.g.,][]{LCB97}.  
Given these problems, the uncertainties in the absolute stellar population ages are still
quite large.  Nonetheless, the observed colors for the majority
 of the isolated dwarf irregular galaxies appear to be most consistent with nearly constant
star formation rates for at least 10 Gyr (Figure \ref{fig:ubv}).  That is, despite their blue
colors, {\em dwarf irregular galaxies are not young systems.}  

 The type of analysis presented here is similar to that presented in \citet{vZ97b} for 
LSB dwarf galaxies.  The models used in \citet{vZ97b} were slightly older versions of the 
\citeauthor{BC96} models, and the same offset existed between the observed and 
theoretical colors.  In that paper, the observed colors were interpreted in the 
context of an aging stellar population, which indicated that the dominant stellar population
in the LSB dwarf galaxies formed at least 1--3 Gyr ago.   The optical colors of the LSB 
dwarf galaxies are actually quite similar to the global colors  of the galaxies in the 
present sample, and thus, in retrospect, it is likely that the LSB dwarf galaxies also
have had approximately constant star formation rates for roughly a Hubble time.
Since the dominant stellar population in a system with continuous star formation activity
will be {\it de facto} the most recently formed stars, i.e., those formed within the 
last few Gyr, this interpretation is consistent with that presented in \citet{vZ97b} for
the LSB dwarf galaxies, but has additional
 implications about the age of the underlying stellar population.

Only the simplest star formation histories, systems where the star formation rate varies
monotonically with time, are illustrated in Figure \ref{fig:ubv}.  However, it is also necessary
to consider that the possibility that the isolated dwarf irregular galaxies have had a more episodic 
star formation history [as has been shown for nearby galaxies like the LMC \citep{vdB91,G95} and 
Carina \citep{SSHL94,M97}].
An example of one of these more complicated star formation histories is illustrated in 
Figure \ref{fig:burstmodels}a, where the locus of points for the dI sample are compared
with the color evolution of a galaxy which
undergoes a brief burst of star formation.   From this representative model, it is clear
that galaxies which are currently undergoing a starburst  have colors
which are too blue for the majority of the isolated dwarf galaxies, and  post--starburst
galaxies would be too red in \ub~to fit the colors of the typical isolated dwarf irregular
galaxy.  Given that the current star formation rates of the isolated dwarf galaxies are
quite low ($\sim$ 0.02 M$_{\odot}$ yr$^{-1}$) it is not too surprising that the starburst 
and post--starburst models do not apply to the majority of the galaxies in this sample.

There are a handful of galaxies which lie outside the locus shown in 
Figure \ref{fig:burstmodels}a, however. UGC 9128, UGC 5205, and DDO 210 are all redder in 
\ub~than expected for a constant star formation rate model, and all have extremely low or 
undetectable current star formation rates.  Both UGC 9128 and DDO 210 have 
color gradients from the inner (blue) to the outer (red) regions.
These galaxies may be examples of post--burst systems. 
 
Two of these three galaxies have more detailed star formation histories derived from
ground--based color--magnitude diagrams.  DDO 210 is a nearby galaxy, and subtends
a large angular size. Ground--based color-magnitude diagrams for DDO 210 indicate
that it has had a recent enhancement of its star formation rate in the central 
regions within the last several Myrs \citep{L99}, but its current star 
formation rate is negligible.  Similarly,
ground-based color-magnitude diagrams for the inner regions of UGC 9128 (DDO 187) 
indicate that its current star formation rate is $\sim$3 times smaller than its maximum

\psfig{figure=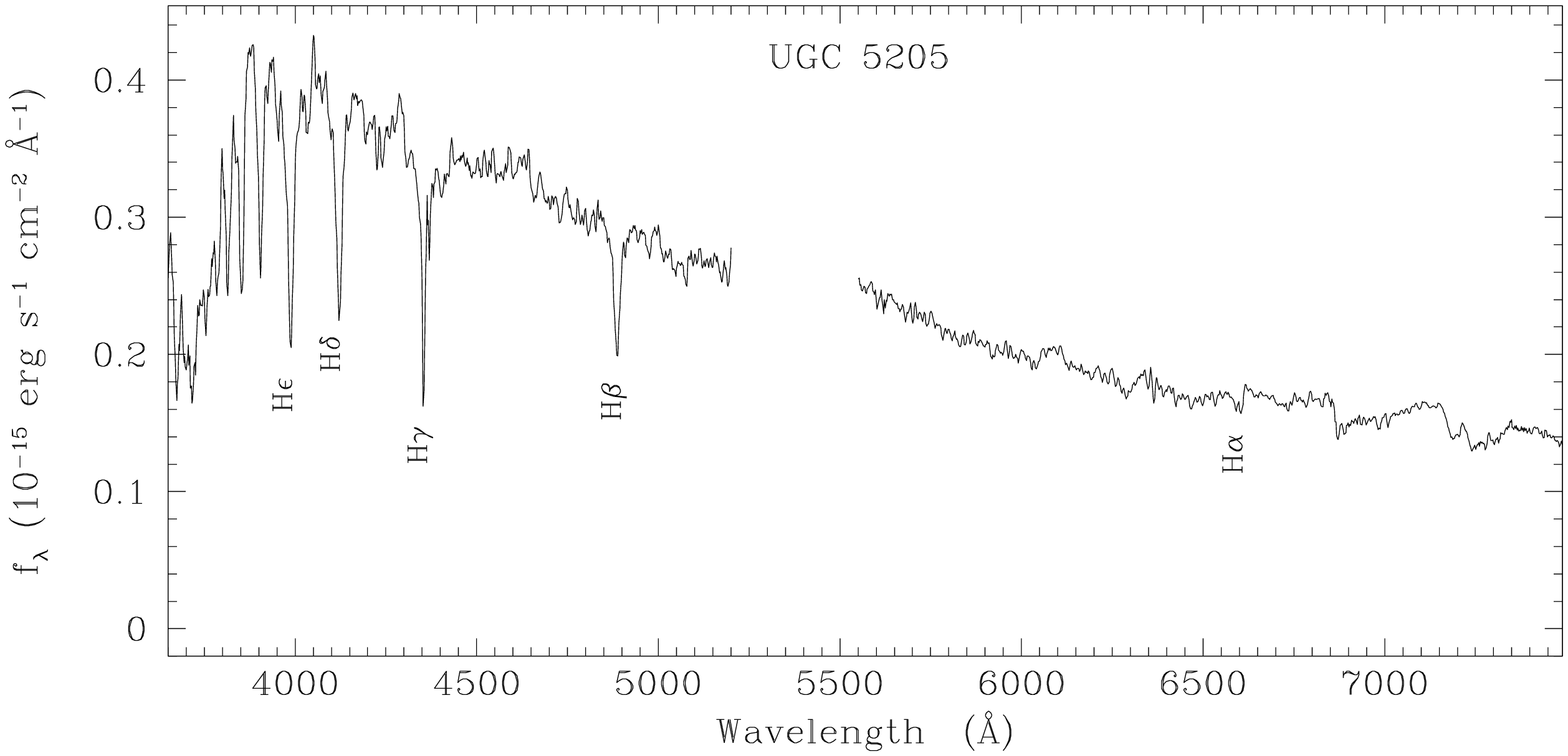,width=9.cm,angle=270.}
\vskip -30pt
\figcaption{Optical spectrum of UGC 5205.  The blue continuum level and presence
of strong Balmer absorption features indicate that the last major episode of
star formation occurred $\sim$1 Gyr ago in this system. \vskip 10pt \label{fig:u5205}}

\noindent star formation rate which occurred a few tens of Myrs ago \citep{ATK00}.  Thus, 
both their global colors and detailed analysis of their stellar populations 
indicate that these galaxies are post--burst systems, in the sense that their 
current star formation rates are substantially lower than those of the recent past.

An accurate star formation history of the third galaxy with anomalous \ub~colors,
UGC 5205, has not yet been obtained, but additional insight into the star formation
history of this galaxy can be gleaned from an optical spectrum of the inner region 
(Figure \ref{fig:u5205}).  While UGC 5205 has no detectable H$\alpha$ emission 
in the narrow band image, the optical colors are still relatively blue; in addition, the inner 
region is knotty and irregular in appearance, suggesting recent star formation activity.
The optical spectrum confirms that UGC 5205 is a post--burst galaxy.  The strong 
Balmer absorption lines combined with the lack of Ca H and K and Mg Ib and
the presence of a blue underlying continuum indicates that the stellar continuum
is dominated by A--stars, exactly as expected for a post--burst system.  Based on
the integrated optical colors and optical spectrum, UGC 5205 appears to have had
its last major star formation episode approximately 1 Gyr ago.   UGC 5205
has a near neighbor (despite being included in an ``isolated dwarf galaxy sample''), 
CGCG 007-025; the velocity difference between these two galaxies is 84 km s$^{-1}$, and the
projected separation is approximately 8.3 kpc.
 Interestingly, the neighboring galaxy has a very high current
star formation rate, and extremely blue colors (it has the bluest \ub~ of the
entire sample). The dynamics of this pair are currently unconstrained, but the
 optical images reveal that UGC 5205  has a 
tidal tail, and thus has been  disturbed by the interaction.  In addition,
the fact that the two galaxies are currently experiencing either an elevated (CGCG 007-025) 
or a quenched (UGC 5205) star formation rate suggests that their interaction
has had a significant impact on the gas distribution and the star formation capability
of the two systems. 

In addition to UGC 9128, UGC 5205, and DDO 210, one other galaxy, UGC 12613, lies outside of 
the general locus of points for the isolated dwarf galaxy sample.  UGC 12613 is the reddest
galaxy in the sample in both \ub~and \bv;  the optical colors are too red to be fit by a 
constant star formation rate model, but can be fit by either a declining star formation
rate model (tau model), or by an aging burst population.  The current star 
formation rate in UGC 12613 is extremely low, which supports the idea that UGC 12613
is dominated by an aging stellar population rather than by one which is constantly being
replenished.  However, there is one fundamental difference between UGC 12613 and the other
three galaxies which may be post--burst galaxies: UGC 12613 is gas--poor.  UGC 12613
has one of the lowest gas mass fractions (0.17) and one of the lowest values of M$_{\rm HI}$/L$_{\rm B}$
(0.34) of all the galaxies in the sample.  Thus, the current star formation rate may be
low as a direct result of the fact that UGC 12613 is in the process of running out of fuel.
The evolutionary tracks in Figure \ref{fig:burstmodels}a indicate 
that if UGC 12613 is a post--burst galaxy, the last major star formation episode was several 
Gyr ago.    This result is consistent with the more detailed stellar population models
derived from observations of the resolved stellar population \citep{Ge98}.  Due to the
small field of view of WFPC2, the HST observations only sampled the southwest quadrant of the galaxy
(which includes the regions of current star formation activity).  \citet{Ge98} present several
possible star formation histories for UGC 12613; their simplest model is that the star
formation rate has been relatively constant for several Gyr, but that there was a recent episode
of enhanced star formation activity which ended approximately 2 Gyr ago.

As demonstrated by these four unusual galaxies, the UBV color--color diagram
can indeed identify systems with ``bursty'' star formation histories.
The power of the star formation history analysis presented here lies in the
fact that the majority of dwarf irregular galaxies lie within a small locus
on the UBV color--color diagram:  the ``typical'' dwarf irregular galaxy has had
an approximately constant star formation rate, and galaxies which undergo a burst
 of star formation are the exception rather than the rule.  This results agrees with 
the seminal work of \citet{HG85,HG86}, based on the integrated colors of a diverse sample
of dwarf and giant Irregular galaxies.  In addition,  the lack of significant
color gradients in the ``typical'' dwarf irregular galaxies (Figure \ref{fig:grads})
 implies that the entire stellar disk must have experienced a
similar star formation history.

Inspection of the H$\alpha$ images presented in \citeauthor{vZ00a} 
\citep[and those in e.g.,][]{HHG93,MH94,M96}
indicate that the current star formation activity is distributed sparsely across the optical
disk of each galaxy.  Thus,  while the optical colors indicate that the star formation activity 
is continuous, it is clear that the local star formation rate fluctuates, presumably as 
dense molecular clouds form and disperse.   As first discussed by \citet{HG85}, one way to
reconcile the local and global star formation histories
is to hypothesize that the star formation activity percolates across the entire
stellar disk in a quasi--continuous manner.
In other words, the global colors indicate a nearly constant star formation rate 
because they represent a temporal and spatial average of
the composite stellar populations.  In this model, 
the ensemble average of the star formation rate for the entire galaxy
is approximately constant and every location in the disk has the potential to be a
site for future star formation activity.  Thus, in contrast to spiral galaxies
where the star formation occurs preferentially in the spiral arms and is presumably
linked to the gas density enhancements from the spiral density waves, {\em star formation
in dwarf irregular galaxies is a purely stochastic process, and has no favored
location within the optical disk.}

\subsubsection{Starbursting Dwarf Galaxies}

Only three of the galaxies in the isolated dI sample were classified as starbursting
dwarf galaxies: UGCA 439, UGC 11755, and CGCG 007-025 all have moderately high star
formation rates, concentrated in the central regions of the galaxy.  Of these three,
only CGCG 007--025 stands out in the UBV color--color diagram as a galaxy with an
unusual \ub~color; UGC 11755 and UGCA 439 blend into the general trend.  However,
these three galaxies do stand out from the general sample in that their outer isophotes
are significantly redder than their inner regions.  Six other galaxies also had
unusual color gradients from the inner to outer regions: UGC 685, UGC 1104, UGC 5288, 
UGC 9240, UGC 10351, and UGC 12713 (Figure \ref{fig:grads}). 
 While these galaxies were not classified as Blue Compact Dwarfs (BCD),
they all have unusually short scale lengths for their luminosities, and may be related
to the BCD class.  As with those galaxies classified as BCDs, the global colors of these 6
``compact dI's'' are not significantly different than those of a typical isolated dI.

This leads to the question of whether the UBV color--color diagram is
sufficient to identify galaxies undergoing a
current burst of star formation if only global colors are used.  As mentioned in Section \ref{sec:col},
the global color is a luminosity weighted color, and may not be representative of the
composite stellar population if there are significant color gradients. The extent to 
which the optical colors are affected will depend not only on 
the strength and duration of the burst, but also on the spatial distribution of the 
young (bursting) stellar population:
a centrally concentrated starburst will only affect the colors of the inner regions,
and may not be sufficient to substantially change the global colors.  

To investigate this further, the UBV colors for blue amorphous dwarf galaxies 
from \citet{MMHS97,MMH99} are shown in Figure \ref{fig:burstmodels}a.  Like the 
starbursting dwarf galaxies in
the isolated dI sample, all of the blue amorphous dwarf galaxies
have a significant color gradient between their inner (starbursting) and outer regions. 
The location of the inner core region and outer envelopes are labelled as ``C'' (core) and ``E''
(envelope) in Figure \ref{fig:burstmodels}a.  As expected for galaxies with centrally
concentrated starburst, the inner core regions are extremely blue while the outer 
envelope regions are much redder than the typical isolated dI.   Clearly, the inner core 
regions of these galaxies are consistent with a starburst population superposed on an 
older (but still blue) stellar population, similar to the burst model shown here
 \citep[also as discussed in detail by][]{MMH99}.  The red outer envelopes are 
consistent with aging stellar populations, 
i.e., ones in which the past star formation rate was higher than the present rate.  
The global colors are also shown in Figure \ref{fig:burstmodels}a (labelled as ``A'' 
for average colors).  Four of the starbursting dwarf galaxies have globally averaged
colors which are typical of dI's, and thus have similar properties to
the three starbursting galaxies found in the isolated dI sample.  The fact that these galaxies have
global colors which are consistent with quasi--continuous star formation activity
indicates that the ensemble average of the inner (young) region and outer (old) region
 yields an intermediate age for the dominant stellar population, despite the 
current burst of star formation.  Nonetheless, it is important to note that these
galaxies are still easily separated from the ``typical'' dwarf irregular because
of their strong color gradients.  Of the remaining galaxies
in the starbursting dwarf galaxy sample, two others (NGC 3125 and NGC 1705) have 
average colors which were similar to their inner core regions, indicating that the 
young stellar population is dominant, and one (NGC 3955) has a global color
 which is as red as the outer envelopes of the other galaxies, indicating that the 
central starburst is only a minor fraction of the total stellar population.

Based on this small sample of starbursting dwarf galaxies, it appears that a spatially resolved star
formation history is extremely important in distinguishing between starbursting dwarf
galaxies and ``typical'' dIs.  Both types of galaxies may have similar global
colors (depending on the relative strength of the star burst), but  the
star formation activity is centrally concentrated in the starbursting dwarf galaxies.
Thus, the result of a starburst is to create a significant color gradient between the inner and
outer regions of the optical disk.

\subsection{Other Constraints on the Star Formation History}
\label{sec:life}

 The next question that needs to be addressed is whether a quasi--continuous star 
formation history is self--consistent with the observed luminosities and 
current star formation rates.  There are two ways of looking at this
question: (1) is the current star formation rate representative? and (2) can
the galaxy continue to form stars at such a rate without running out of fuel
in the near future?  The first question addresses the issue of whether the
star formation rate is approximately constant as a function of time while
the second question addresses the
issue of whether the fuel supply is sufficient to support the current
star formation rate.

\subsubsection{Average Past Star Formation Rate}

The average past star formation rate can be derived by estimating the total mass of 
stars formed and dividing this quantity by the length of time for which star 
formation has been taking place:
\begin{equation}
   <SFR>_{\rm past}~  =  {M_{\rm stars} \over T_{\rm sf} }.
\end{equation}
The latter quantity is relatively unconstrained, 
but, at maximum, T$_{\rm sf}$ must be less than the age of the universe.  
The total mass of stars formed is also uncertain, but can be estimated if one assumes a stellar 
mass--to--light ratio ($\Gamma_b$) and estimates the amount of
 material that has been recycled during the history of the system ($R$):
\begin{equation}
    M_{\rm stars} = {\Gamma_b~ L_B \over (1-R)}. 
\end{equation}
\citet{KTC94} investigated the average past star formation rates for a large 
sample of galaxies, spanning a range of Hubble 
types.  They adopted a type--dependent mass--to--light ratio, a uniform
recycling fraction of $R = 0.4,$ and a T$_{\rm sf}$ of 10 Gyr.  They found 
that, on average, the earlier Hubble types had \citet{S86} $b$ parameters 
(ratio of current--to--past star formation rates) that were lower than later
Hubble types, but that spiral galaxies generally had lower current star formation
rates than their average past rate.  This result is consistent with the idea
that spiral galaxies have decreasing star formation rates, and that the 
observed color differences between the 
Hubble types may be a result of different star formation histories 
\citep[e.g.,][]{T68,SSB73}.

Adopting a similar formalism, it is possible to derive the ratio of the
current--to--past star formation rates for the isolated dwarf Irregular galaxy sample.
One of the difficulties for this type of analysis is that the stellar mass--to--light
ratio is highly uncertain.  The stellar mass--to--light ratios adopted by \citet{KTC94} were
based on dynamical mass estimates, corrected for the contribution of dark matter;
they adopted a V--band mass--to--light ratio of 0.6 for the irregular galaxies in their sample.
However, the dynamical mass--to--light ratio is not well constrained for irregular 
galaxies; most dwarf irregular galaxies have solid body rotation throughout the optical
disk \citep[e.g.,][]{SBMW87,MCR94,vZ97c}, and solid body rotation curves cannot be well 
fit by maximum disk models.  
In a similar analysis of the star formation history of dwarf galaxies,
\citet{MH94} adopted a color dependent conversion formula for the mass--to--light ratio 
based on galaxy evolution models of \citet{LT78}. In their study, a typical dwarf 
galaxy (\bv~= 0.4) has a mass--to--light ratio of 1.0 in the B--band.

While it is somewhat circular reasoning to use a galaxy evolution code to derive a
color dependent mass--to--light ratio in order to check the galaxy evolution model,
at the very least such a model will provide a self--consistent estimate of the stellar mass. 
Thus, for consistency with the previous discussion, the mass--to--light ratio was derived 
using the \citeauthor{BC96} code and a quasi--continuous star formation rate model:   
\begin{equation}
       {\rm log~} \Gamma_b= 2.84 (B-V) -1.26 
\end{equation}
where \bv~is the observed color. This formalism results in a typical $\Gamma_b$ of 0.8
for  dwarf irregulars (\bv~of 0.4), and is valid only for galaxies
with \bv~$<$ 0.5.

The second necessary parameter for the derivation of the average past star formation rate
is the recycling fraction, which is can be derived from the ratio between the total mass
of stars formed over the history of the system and those still luminous at
the present epoch.  Using the \citeauthor{BC96} code, the recycling
fraction ($R$) for a quasi--continuous star formation rate was found to be $\sim$ 0.33; that is,
the total mass of stars formed is a factor of 1.5 times more than is presently
detectable. 

The final ingredient for this analysis is the age of the stellar population.  
Current estimates of the onset of star formation activity in the universe are 
on the order of $z\sim 5 - 10$ \citep{SAGDP99,MSSB00}.  Whether the galaxies in this sample 
began forming stars at this time is unknown, but it is not unreasonable to
assume that they have been forming stars for approximately a Hubble time; thus, 
for H$_0$ = 75 \kms Mpc$^{-1}$, T$_{\rm sf} \sim$ 13 Gyr.  

\psfig{figure=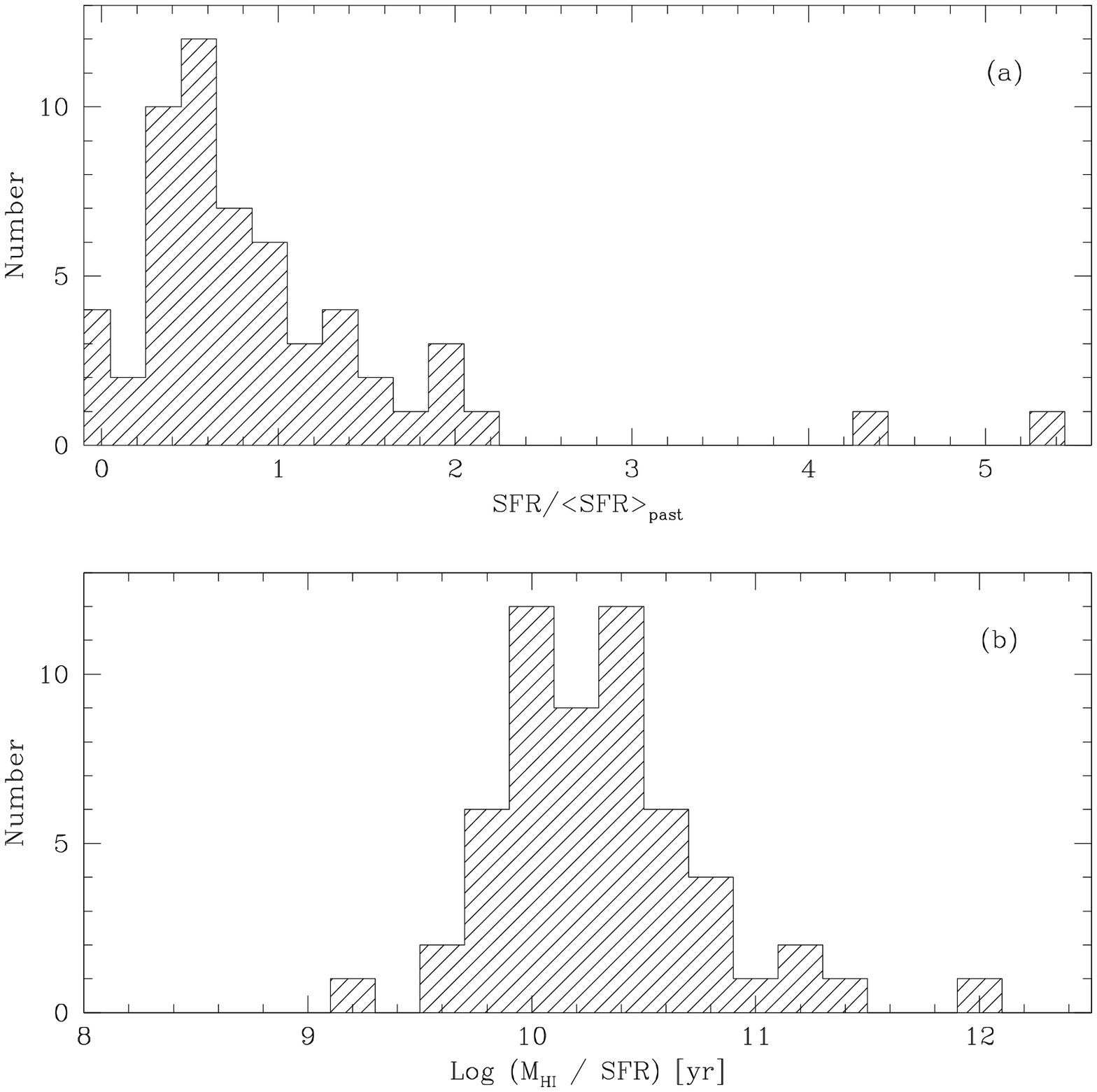,height=8.cm,angle=0,bbllx=20 pt,bblly=170 pt,bburx=700 pt,bbury=700 pt,clip=t}
\figcaption{(a) The current star formation rate normalized by
the average past star formation rate.  For most of the dwarf irregular galaxies, the current
star formation rate is approximately equal to the average past rate, supporting
the idea that star formation is quasi--continuous.  (b) The gas depletion time scales.
Most of the dwarf irregular galaxies have long gas depletion time scales (on order of a
Hubble time or more), and thus can continue to form stars at the present rate 
without running out of fuel. \vskip 10pt \label{fig:life}}

The combination of higher mass--to--light ratios, a longer 
T$_{\rm sf}$, and a
slightly different current SFR conversion factor results in Scalo $b$ parameters
which are a factor of $\sim$0.87 times lower than the formalism for
Irregular galaxies in \citet{KTC94}.   A histogram of $b$ parameters 
derived from the above equations are shown in Figure \ref{fig:life}a; the
median value is 0.7 $\pm ^{0.5}_{0.2}$.  For a quasi--continuous
star formation rate, the current star formation rate should be similar
to the average past rate.  Thus, within the accuracy of this type of calculation,
it appears that a quasi--continuous star
formation rate is consistent with the observed integrated luminosities.
  Note that the extreme outliers ($b > 2$) are
the three star bursting galaxies in the sample, which have much higher current
star formation rates than their average past rates.

In summary, both the optical colors and the integrated luminosities are well
described by a constant star formation rate model.  While it is important
to remember that neither the luminosities nor the optical colors will be sensitive to 
a well evolved burst population (Figure \ref{fig:burstmodels}),
the fact that the current (low) star formation
rates appear to be sufficient to build the observed stellar population over
a Hubble time is compelling.  If episodic bursts are the dominant
mode of star formation, a larger fraction of the galaxies should be
in a post--burst state, and there should be no correlation between the
current star formation rate and the average past star formation rate.
 Thus, while it is impossible to completely rule out the possibility of early 
bursts of star formation, {\em the majority of galaxies
in this sample do not need to have had substantial bursts at any epoch
to build their present luminosity}.  
These simple arguments suggest that the dominant mode of star formation in dwarf irregular
galaxies is one where star formation percolates slowly across the disk, with an
nearly constant average star formation rate.

\subsubsection{Gas Depletion Timescales}
Another important timescale argument concerns whether these galaxies can continue to sustain
their current star formation activity for several more Gyr. The gas depletion
time scale was calculated from the observed neutral hydrogen mass and current
star formation rate:
\begin{equation}
\tau_g={ M_{HI} \over SFR.}
\end{equation}
 A histogram of the gas depletion time scales are shown in Figure \ref{fig:life}b; 
the median gas depletion time scale is 19 $\pm^{14}_{8}$ Gyr,  This clearly indicates
that these galaxies are in no danger of running out of fuel, and
can continue to form stars at the current star formation
rate for at least another Hubble time.

It is important to note, however, that the quoted gas depletion timescales assume 
that the galaxy will be able to convert all of the available gas into stars.
Spatially resolved maps  of the neutral hydrogen distributions 
of dwarf galaxies indicate that a large fraction of the total hydrogen mass is 
located beyond the optical radius \citep[e.g.,][]{SBMW87,STTv88,CB89,LSY93,HSFRWH96,vZ97c}.  
It remains a mystery as to whether this extended gaseous material will be available for 
future star formation activity in these systems \citep[see, e.g.,][]{GH84}.  

The simple lifetime arguments presented here indicate that the
observed colors, luminosities, and gas properties are consistent with
a quasi--continuous star formation rate.  The fact that the isolated dwarf irregular
galaxies have sufficient fuel to sustain similar star formation
activity for at least another Hubble time provides additional support to the
idea that star formation will
continue to percolate (at a low rate) across the disks of these galaxies
in a quasi--continuous manner for many Gyr to come.

\subsection{Comparison to Resolved Stellar Population Models}
\label{sec:cmds}

As mentioned in the Introduction, the preferred method of determining the star formation
history of a galaxy is to observe the resolved stellar population and count the relative
number of stars in various evolutionary phases (main sequence, horizontal branch, blue loop, etc.).
Color--magnitude diagrams (CMDs) provide a direct means of converting the number of
stars at these evolutionary phases into a detailed star formation history.  However, 
most galaxies are too distant to make such an analysis feasible.
  There are a few galaxies in the present sample which have been observed with HST, 
or are near enough that ground--based CMDs have been constructed, but these galaxies happen to 
be non--representative of the sample as a whole.  The results of CMD analysis of UGC 9128, DDO 210, 
and UGC 12613 were presented in Section \ref{sec:sfh}.1, where it
was shown that all three of these galaxies are likely to be post--burst systems, in agreement
with the observed optical colors.  None of the other galaxies in the isolated dI sample have
been observed with sufficient spatial resolution to derive detailed star formation histories, but it is
nonetheless instructive to compare the results from integrated optical colors with the
star formation histories derived from the resolved stellar populations of the nearest
 dwarf irregular galaxies.

Observations of the resolved stellar populations in starbursting dwarf galaxies indicate that star formation 
activity is both temporally and spatially confined. For example,
the current star formation activity dominates the central regions of VII Zw 403 (UGC 6456), 
one of the nearest blue compact dwarf galaxies \citep{LTNH98}.  Analysis of the resolved stellar population
indicates that the star formation rate peaked a few 100 Myr ago in VII Zw 403, and that the star formation activity
is in a more quiescent stage at the present epoch \citep{LTNH98,SHCG99}.  At its peak, the star 
formation rate was $\sim$ 30 times higher than the typical star formation rate; this ``burst'' phase
lasted for a few 100 Myr before returning a low level of star formation activity.  \citet{SHCG99} emphasize
that this was not the first episode of star formation in VII Zw 403, since intermediate--age and old metal--poor
stars are distributed throughout the galaxy.  

In sharp contrast with the starbursting dwarf galaxies, the ``normal'' dwarf irregular
galaxies appear to have a more smoothly varying star formation history.
 \citet{DP98} summarized the results of HST observations of the resolved stellar 
populations of four Local Group dwarf galaxies: Sextans A, GR 8, UGC 12613 (Peg DIG), and Leo A.   
Not unexpectedly, each galaxy in this sample has a unique star formation history; however, certain
commonalities exist.  Of these four, only Sextans A
appears to have had significant variations in the star formation activity over the last Gyr;
however, unlike VII Zw 403, the current elevated star formation rate in Sextans A is only a 
factor of 3 higher than the past rate.  In other words, none of these four galaxies appear to have 
undergone a significant starburst episode within the last Gyr.  
 Thus, analysis of the resolved stellar
populations provides additional confirmation that the star formation activity in most
dwarf irregular galaxies occurs in a quasi--continuous manner, with a slow, but steady,
conversion of gas into stars.

 Finally, based on the spatial distribution of the resolved stellar population, \citet{DP98} found
 that star formation occurs at a low level across the entire stellar disk.  In all four dI galaxies, 
the localized star forming complexes were found to occur on spatial scales similar to those of giant molecular 
clouds ($\sim$ 100 pc), and to last $\sim$ 100 Myr. 
At this time, only a few gas--rich dwarf galaxies have been observed with sufficient spatial resolution and
depth to derive spatially resolved star formation histories.  Clearly, further observations of the 
nearest gas--rich dwarf galaxies are needed to investigate the propagation of star formation activity
in small galaxies.  Of particular interest is the question of whether subsequent star
formation activity is predominately a stochastic process, or if star formation propagates from one site
to another in a more organized manner (as predicted by SSPSF).

\section{Current Star Formation Activity}

As discussed in Section \ref{sec:life}.1, the current star formation rate of the dI sample
is approximately
the same as the average past star formation rate.  This implies that there is a correlation
between current star formation rate and luminosity, since more luminous galaxies must have had a 
higher star formation rate in the past.  It is thus necessary to normalize the current star formation
rate by a scaling factor to remove  obvious scaling correlations (i.e., big 

\psfig{figure=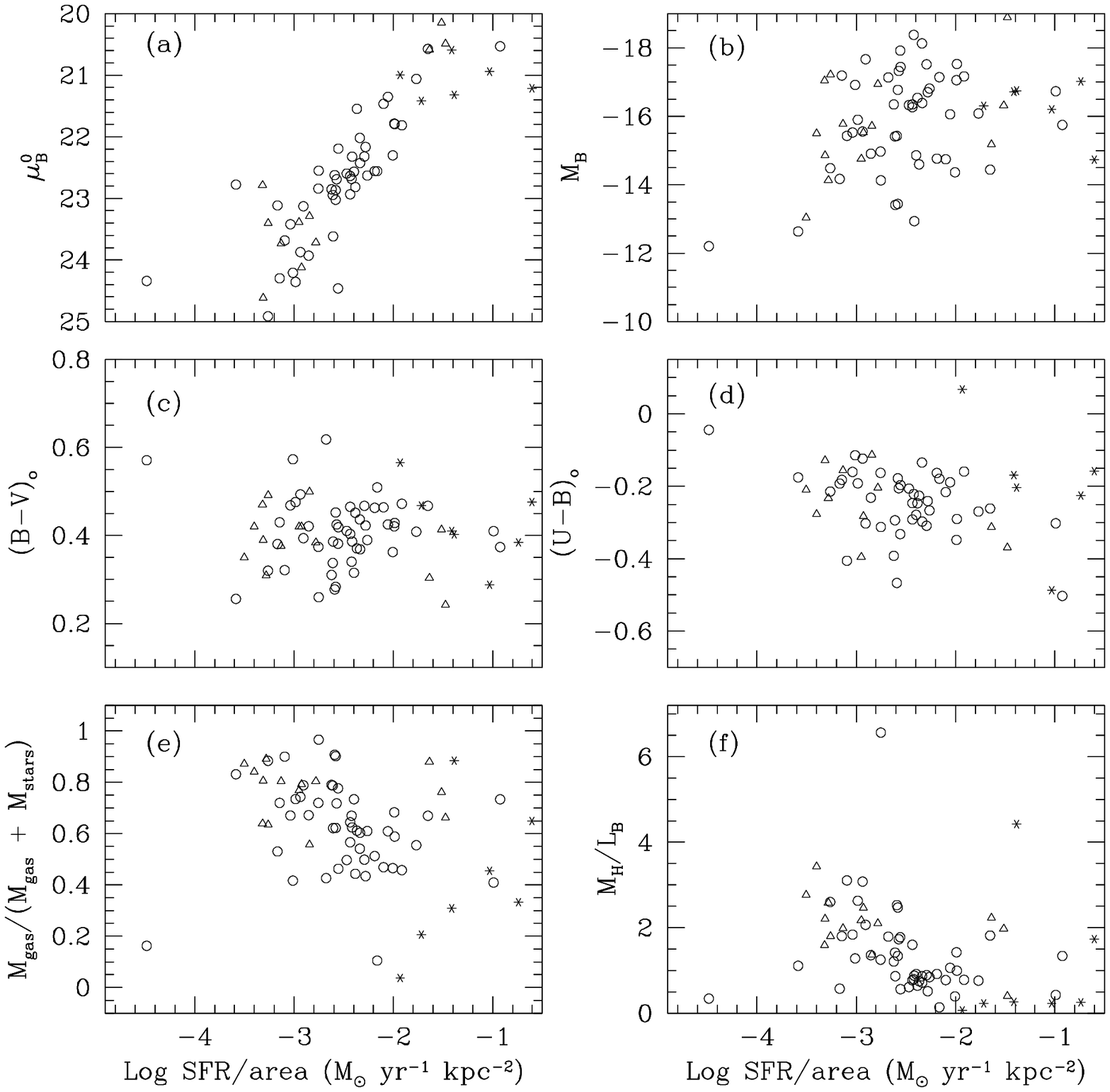,height=8.cm,angle=0,bbllx=20 pt,bblly=160 pt,bburx=700 pt,bbury=700 pt,clip=t}
\figcaption{Correlations between the current star formation rate and other physical
properties. Galaxies from the isolated dI sample are denoted by open circles; galaxies
from van Zee et al.\ (1997) are denoted by open triangles; blue amorphous dwarf galaxies from
Marlowe et al.\ (1997) are denoted by asterisks.
 The star formation rate has been normalized by the size of the galaxy,
where area = 2$\pi$(1.5$\alpha$)$^2$.
(a) A significant trend is evident in the relation between star formation
rate per unit area and central surface brightness.  Galaxies with low central surface
brightness also have low normalized star formation rates.  
(b) No correlation is found between
absolute magnitude and the normalized star formation rate.
(c) No correlation is found between normalized star formation 
rate and \bv~or 
(d) \ub.
(e) Correlation between
gas mass fraction (baryonic) and star formation rate.  In this sample, galaxies which are currently
gas--rich also have low normalized star formation rates, although the converse does not hold.   
(f) A weak correlation is seen between gas--richness, as measured by M$_H$/L$_B$, and star 
formation rate per unit area.  \vskip 10pt
\label{fig:sfr}}

\noindent   galaxies are big).
One way to normalize the current star formation rate is to divide by the
surface area of the galaxy \citep[e.g.,][]{HG86}.  In the following analysis, ``area''
has been defined in terms of scale length:   $r$=1.5$\alpha$. 
 The choice of 1.5$\alpha$ is  arbitrary, but it roughly corresponds to $R_{26.5}$ for this sample.
The normalized current star formation rate is plotted against a variety of physical parameters in 
Figure \ref{fig:sfr}; these plots include galaxies from both the present sample,
the LSB dwarf galaxies from \citet{vZ97b}, and the blue amorphous dwarf galaxies from \citet{MMHS97}.
 The use of a normalized star formation rate removes
the correlation between star formation activity and luminosity (Figure \ref{fig:sfr}b) but
introduces a correlation between surface brightness and normalized star formation rate (Figure
\ref{fig:sfr}a).  However, this is exactly the correlation one would expect for galaxies
with quasi--continuous star formation:  high surface brightness galaxies must have had
 high star formation rates per unit area \citep[see also][]{H97}.
Thus, in the subsequent plots one can consider either surface brightness or normalized
star formation rate as the ordinate on the abscissa.  

Figure \ref{fig:sfr}(c,d) shows that there is no correlation between normalized star formation
rate and the optical colors.  That is, the normalized star formation rate (and, consequently, the
surface brightness) is independent of the exact star formation history.  This lack of correlation
between age and surface brightness is also seen in large LSB disk galaxies  \citep[e.g.,][]{OBSCI97,Be99,K99}.
Thus, the central surface brightness of a galaxy appears to be an intrinsic property, not a  
result of differing evolutionary stages.

The final two plots in Figure \ref{fig:sfr} investigate the relationship between current star formation
activity and gas properties of dwarf galaxies.  The gas properties are particularly relevant
because the gas provides  fuel for star formation activity.  In many ways, the gas--richness
of a galaxy is a good indicator of evolutionary status, insofar as a gas--rich galaxy has not converted much
of its gas into stars.  Note, however, that this does not necessarily imply that a gas--rich galaxy
is ``young,'' since a galaxy with a low star formation rate will take longer to convert
gas into stars.  

Figure \ref{fig:sfr}e shows a weak trend of normalized star formation rate
with gas mass fraction.  The gas mass fraction is the ratio of the total
gas mass (M$_{\rm gas}$) to the total baryonic mass 
(M$_{\rm gas}$ + M$_{\rm stars}$).  The stellar mass
was computed as described in Section \ref{sec:life}.1 and the gas mass includes
a correction for the neutral helium component [M$_{\rm gas}$ = 1.3 $\times$ M$_{\rm HI}$, \citep{AG89}].
The molecular gas component is neglected for this computation since it is difficult to 
detect CO in low mass galaxies \citep[e.g.,][]{EEM80,TKS98}, and, further, the correction from CO 
to H$_2$ column density is highly uncertain for low metallicity galaxies
 \citep[e.g.,][]{VH95,W95}.  For a simple model of 
galaxy evolution, the gas mass fraction is initially 
unity (pure gas disk) and then decreases as the gas is converted into stars.  
  The correlation seen in Figure \ref{fig:sfr}e indicates that
higher surface brightness galaxies (high normalized star formation rate galaxies) 
have lower gas mass fractions. This is exactly the trend expected if the average past
star formation rate is similar to the present rate: galaxies with high star formation
rates will convert a larger fraction of their gas into stars by the present epoch, 
and therefore have both higher surface brightness and lower gas mass fractions.
 This trend is also seen in samples of disk galaxies \citep{MdB97}. 
 However, there are a few galaxies which counter this trend, such
as UGC 12613.  UGC 12613 has both a low normalized star formation rate and a 
low gas mass fraction.  In fact, dwarf transition galaxies \citep[e.g.,][]{KSG99} and
 gas--poor ellipticals, such as NGC 205 \citep{YL97},
should fill in the lower left region of the diagram.  
The upper right, however, is likely to be 
genuinely devoid of galaxies at the
present epoch since a high normalized star formation rate will rapidly process the
gas, and thus a galaxy should evolve quickly from this region.

A similarly weak trend in seen in the plot of M$_{\rm HI}$/L$_{B}$ (Figure \ref{fig:sfr}f).
The galaxies in this sample span only a small range in 
M$_{\rm HI}$/L$_{B}$, but the gas--rich systems again appear to be the systems with 
lower normalized star formation rates (again with the exception of UGC 12613).  
However, it is also important to note that not all low surface brightness galaxies are
unusually gas--rich, and, conversely, not all gas--rich galaxies have extremely 
low surface brightness \citep[e.g., DDO 154,][]{CB89}.

The majority of the galaxies in the present sample are gas--rich systems, with gas mass
fractions greater than 0.4 (Figure \ref{fig:sfr}e).   However, if these galaxies have
an abundant supply of gas, why haven't they converted it into stars in an efficient manner?
In fact, one of the puzzling results of studies of the gas distribution in 
dwarf galaxies is that both extreme LSB and  ``normal'' dwarf irregular 
galaxies have gas densities which are well below the threshold for star formation 
activity, despite the fact that these galaxies are considered gas--rich systems 
\citep[e.g.][]{HP96,vZ97c,HEB98}.  This result may hold the key to explaining both
why these galaxies are still gas--rich and why the star formation activity is
quasi--continuous in normal dwarf irregular galaxies: the gas disk is globally stable
against gravitational collapse.  If this is correct, star formation can occur only in 
localized regions of enhanced gas density.  Further, since the likelihood is small
that any region will increase its density enough to exceed the threshold density,
 star formation activity will occur at a very low rate across the stellar disk.   
Combined with the fact that the gas density is almost constant throughout the optical extent,
this suggests that star formation activity percolates randomly across the entire stellar disk
in an inefficient manner, thereby creating a dominant young stellar population at all 
locations while retaining a high gas mass fraction.

\section{Conclusions}

Optical colors and global star formation rates for a sample of isolated dwarf
irregular galaxies have been interpreted in the context of composite stellar
population models.  The major results are summarized below.

(1) The majority of dwarf irregular galaxies have minor or non--existent
color gradients across their stellar disks, and thus global colors accurately
reflect the evolutionary status of the entire stellar component.  The few 
galaxies in this sample which do have significant color gradients
have very short scale lengths for their luminosities, and may be related to the 
class of starbursting dwarf galaxies.

(2) The global colors are blue, with median values of 
\bv~= 0.42 $\pm ^{0.04}_{0.05}$ and \ub~= --0.22 $\pm ^{0.04}_{0.07}$.
 The observed luminosities, optical colors, and current star formation rates
are consistent with an approximately constant integrated
star formation rate over the age of the universe \citep[see also][]{HG85}.  In addition,  typical
gas depletion timescales are $\sim$ 20 Gyr, indicating that these galaxies can
continue to form stars at the present rate for at least another Hubble time.

(3) There is no correlation between surface brightness and star formation history.
The only significant correlation in physical parameters 
is that low surface brightness galaxies have low normalized star formation rates
\citep[see also][]{HG86}.  Weak trends suggest that most low surface brightness
galaxies are also gas--rich, as traced by gas mass fraction and M$_{\rm HI}$/L$_{\rm B}$.

At the very minimum, galaxy evolution models must be able to reproduce the global properties
of a galaxy such as colors, luminosity, current star formation rate, gas mass fraction, and
metallicity.  As shown throughout this paper, the observed global properties of the
typical dwarf irregular galaxy are well fit by a constant average star formation rate, 
with a typical age of $\sim$ 10 Gyr.  Combined with the studies of the resolved stellar
populations of dwarf galaxies in the Local Group \citep[e.g.,][]{DP98}, this argues
against the possibility that most dwarf irregular galaxies have a significant starburst
phase.  Rather, star formation percolates throughout the stellar disk at a slow, but 
approximately constant, rate.  Thus, while starburst models are necessary to explain
the observed properties of a few dwarf galaxies, these galaxies are the exception, not
the rule.

If this interpretation is correct, then the majority of dwarf irregular galaxies
in the local universe are unlikely to be the evolutionary remnants of the
``faint blue galaxies'' found at intermediate redshift \citep[e.g.,][]{CSH91,DWG95}. 
Several models have shown that it is possible to reproduce the excess of faint
blue galaxies found in deep surveys by including a rapidly evolving population
of low luminosity galaxies whose star formation rates peak at $z \sim 1$
\citep[e.g.,][]{BR92,BF96}.  The basic reasoning of these models is that the luminosity
of a starbursting dwarf galaxy may be sufficiently elevated to allow it to be included
in a deep survey, but its evolutionary remnant will be too faint to be included in
surveys of the local galaxy population.  Thus, the difference in the derived
 local and intermediate-z luminosity functions creates 
an ``excess'' of faint blue galaxies at intermediate--z.  However, these models
require that formation of dwarf galaxies be delayed until $z \sim 1$ so
that the star formation rates peak at the appropriate epoch.  This does not
appear to be the case for the typical dwarf irregular, and thus the dwarf irregular class
probably does not contribute significantly to the excess population of faint blue galaxies.
Of course, these models do not rule out the possibility that the faint blue galaxies are
related to gas--poor objects at the present epoch, such as dwarf elliptical galaxies.

Finally, the star formation history of the typical dwarf irregular galaxy
is substantially different than that of a  massive spiral or elliptical galaxy.
Standard stellar population models for spiral galaxies indicate that the observed
colors are best fit by declining star formation rates \citep[e.g.,][]{LT78} while
the colors of elliptical galaxies are best fit by evolving stellar populations
\citep[e.g.,][]{W94}.  Schematically, the  star formation history of the
universe is also in agreement with these stellar population models; the most
recent models indicate that the star formation rate of the universe was substantially
higher in the past than it is today \citep[e.g.,][]{PF95,LTHCL95,Me96}.  
In contrast, dwarf irregular galaxies appear to be forming stars 
at approximately the same low rate now as they were several Gyr ago.
Further, since dwarf galaxies are still extremely gas--rich systems, they can continue to
convert gas into stars at the same low rate for many Gyr to come.
Thus, at some point in the future, dwarf irregular galaxies may be the only systems
which will retain sufficient fuel to continue forming stars; in other words,
it is only a matter of time before dwarf 
galaxies  dominate the star formation activity of the
universe.

\acknowledgements
Elizabeth Barton, David Schade, and Bev Oke are gratefully acknowledged for numerous
thought--provoking conversations about galaxy formation and evolution.
St\'ephanie C\^ot\'e and Vicki Weafer provided helpful comments on early versions of
this paper.
Martha Haynes kindly provided access to the Palomar 5m telescope for the
spectroscopic observations of UGC 5205.  
This research has made use of the NASA/IPAC Extragalactic Database (NED)   
which is operated by the Jet Propulsion Laboratory, California Institute   
of Technology, under contract with the National Aeronautics and Space      
Administration.

\begin{table}
\dummytable\label{tab:obs}
\end{table}

\begin{table}
\dummytable\label{tab:derived}
\end{table}

\psfig{figure=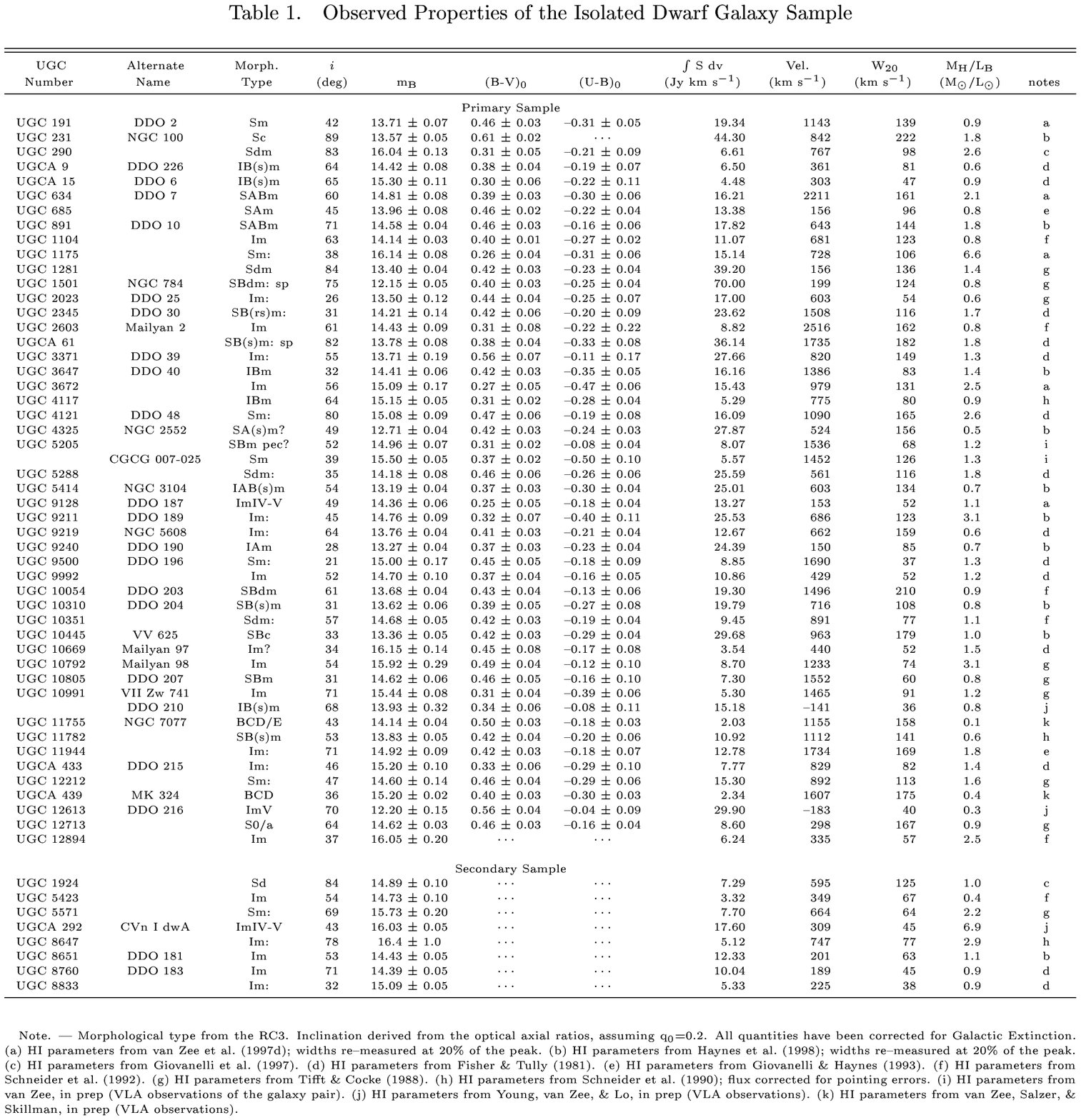,height=25.cm}

\psfig{figure=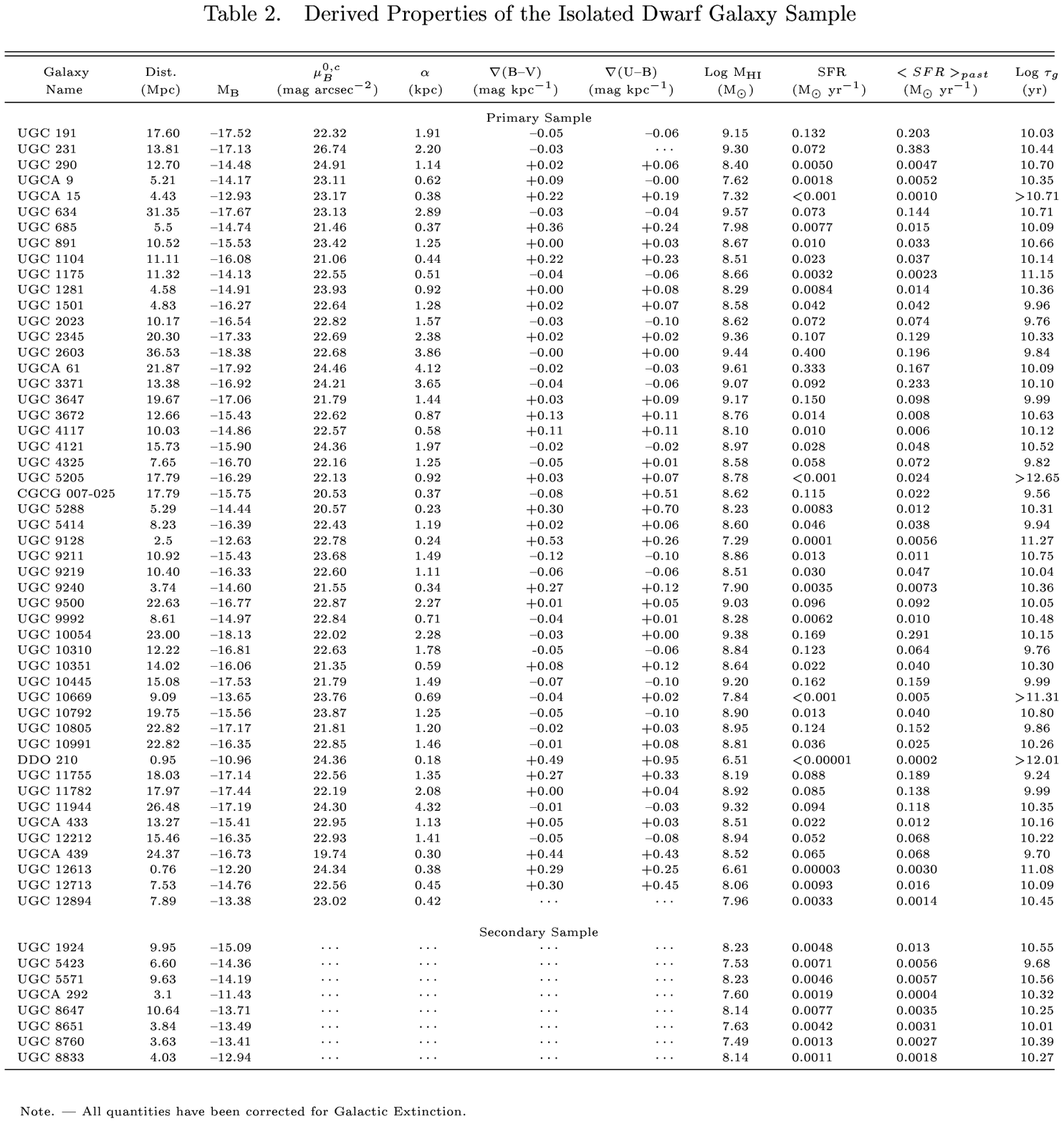,height=25.cm}

\end{document}